# Refining Student Marks based on Enrolled Modules' Assessment Methods using Data Mining Techniques

Mohammed A. Alsuwaiket
Department of Computer Science and Engineering Technology, Hafar Batin University, Hafar Batin, Saudi Arabia
malsuwaiket@uhb.edu.sa

Anas H. Blasi
Department of Computer Information Systems, Mutah University,
Karak, Jordan
ablasi1@mutah.edu.jo

Khawla Altarawneh
Department of Computer Science, Mutah University,
Karak, Jordan
khawla_t@mutah.edu.jo

*Abstract*—Choosing the right and effective way to assess students is one of the most important tasks of higher education. Many studies have shown that students tend to receive higher scores during their studies when assessed by different study methods - which include units that are fully assessed by varying the duration of study or a combination of courses and exams - than by exams alone. Many Educational Data Mining (EDM) studies process data in advance through traditional data extraction, including the data preparation process. In this paper, we propose a different data preparation process by investigating more than 230,000 student records for the preparation of scores. The data have been processed through diverse stages in order to extract a categorical factor through which students' module marks are refined during the data preparation stage. The results of this work show that students' final marks should not be isolated from the nature of the enrolled module's assessment methods. They must rather be investigated thoroughly and considered during EDM's data pre-processing stage. More generally, educational data should not be prepared in the same way normal data are due to the differences in data sources, applications, and error types. The effect of Module Assessment Index (MAI) on the prediction process using Random Forest and Naive Bayes classification techniques were investigated. It was shown that considering MAI as attribute increases the accuracy of predicting students' second year averages based on their first-year averages.

## I. INTRODUCTION

During the last few decades the use of coursework-based module's assessment has increased in the UK and other countries due to various educational arguments justifying it. The students prefer assessments based on coursework alone or a mix of both coursework and exams because these types of assessments tend to yield higher marks than exam based assessment alone [1]. The expanded use of coursework-based evaluation has added in the imprints of singular modules and has led in the extension of good degrees [2]. An expanding number of colleges moved from conventional test based to consistent appraisal all through the semester (for example coursework-based) [3]. Coursework-based evaluation techniques differ from test based appraisal strategies where the learning or the ability is tried in a quite certain timeframe. In addition, it has been generally recognized that the picked appraisal strategy will decide the style and substance of understudy learning and ability procurement [3]. Coursework imprints are a superior indicator of long haul learning than tests [4]. To the best of our knowledge, no studies have considered the increase in marks in the data pre-processing phase. This led to a conclusion that applying conventional DM processes on educational data may not produce accurate results. Authors in [5] proposed a model that refines students' marks based on enrolled modules' assessment methods. The model represents a sub process through which module assessment methods are considered for further processing using a new attribute that reflects the ratio of coursework weightings. Although educational data have been recorded and analyzed from educational software for long, only recently this process has been formed into the new field of Educational Data Mining (EDM) [6]. The EDM process converts raw data from educational systems into useful information that could potentially have a greater impact on educational research and practice [6]. Additionally, EDM uses a wide range of methods to analyze data, including - but not limited to- supervised and unsupervised model induction, parameter estimation, relationship mining, etc. [7, 8].

## II. ASSESSMENT METHODS IN HIGHER EDUCATION

Regardless of the inability to absolutely ensure student learning, assessment is still an essential tool through which teachers have to influence the ways students respond to courses. There are clear steers from UK government towards coursework-based assessment focused on employability [9]. However, not all assessment methods suit all programs or all courses [10]. Thus, student assessment methods in Higher Education can be generally divided into two main categories: Exam-based assessment, which includes different forms of exams such as closed and open book examination, essay-type exams, multiple-choice exams, etc. and coursework-based assessments, which include research projects, assignments, etc. Different studies have proved that students tend to gain higher marks from coursework-based assignments than they do from examinations [11]. It was also found that combined exam-based and coursework-based assessment methods produced up to 12% higher average marks than examinations. Author in [10] conveyed an analysis of marks on more than 1,700 modules at Oxford Polytechnic. Modules with 100%

Corresponding author: Anas H. Blasi

coursework had an average mark 3.5% higher than modules with 100% examination assessment.

This paper solves the problem of a gap between coursework and exam based assessments and tries to normalize students' marks based on the assessment method of the enrolled modules.

### III. STUDENT TRANSCRIPT DATA

In this paper, 4662 modules with different assessment methods were investigated. The basic average calculations of module marks in all departments showed that students who are assessed by coursework tend to get higher marks than those who are assessed by exams or a mixture of both coursework and exams, as shown in Table I.

TABLE I. AVERAGE MODULE MARKS OF STUDENTS IN EACH DEPARTMENT AND THEIR ASSESSMENT METHODS

| Department | Number of students | Average module mark of students by assessment type | | |
|---|---|---|---|---|
| | | Exam-based | Coursework-based | Mixed |
| Business | 54960 | 59.77 | 60.83 | 60.01 |
| Civil Engineering | 34892 | 58.78 | 63.74 | 60.70 |
| Computer Science | 19800 | 58.18 | 64.40 | 58.87 |
| Electrical and Computer Systems Engineering | 13740 | 59.55 | 63.26 | 57.00 |
| Mathematics | 24152 | 61.59 | 66.00 | 61.17 |
| Mechanical Engineering | 31385 | 58.80 | 64.26 | 60.24 |

A simple t-Test was applied to the data of Table I in order to measure the difference between the means of each pair of variables. Results show that there is a statistically significant difference between Exam-based and Coursework-based assessments (with 95% confidence level which equates to declaring statistical significance at the $p<0.05$ level), a t-value of -5.06 and a p-value of 0.001). Applying the t-test to measure the significance of difference between each pair of variables (Ex-CW, Ex-Both, and CW-Both assessment methods) has the following results.

- Significance of difference between exam-based and coursework-based assessments: By applying the paired t-test on the above fields, with 95% confidence level (which equates to declaring statistical significance at $p<0.05$), a t-value of -5.83 and a p-value of 0.002 were obtained. Since the p-value is less than 0.05, it can be concluded that there is a statistically significant difference between the Exam-based and Course-based assessments.

- Similarly, the p-value of the paired t-test between the coursework-based and mixed assessments (0.004) was also less than 0.05, which also indicates the statistically significant difference between these two assessment methods.

On the other hand, it seems to be no statistically significant difference between the Exam-based assessment and the mixture of both exam and coursework assessment methods, since the p-value was 0.749 which is greater than 0.05.

#### A. Understanding Student Transcript Data

Student transcript data were collected systematically. The data consisted of files with hundreds of thousands of records. The data in these files had first to be understood, then cleaned, and finally factors that have the most significance had to be highlighted in order to further process these data using EDM techniques. Therefore, prior to modeling the data in either DM or EDM, the data types should be first identified and understood. Normally, educational data are discrete, i.e. either numeric or categorical data, and noise-free. The lack of noise in educational data is due to the fact that they are not measured, they are either collected automatically or checked carefully [12]. On the other hand, missing data values exist, usually in the cases where students skip answering a given questionnaire or when teachers skip checking attendance data. Humans normally do this type of errors, which are generally referenced as data entry errors [13]. The investigated data represent around 230,823 student records representing a total of six departments at a UK University. Each one of these department data sheets contains a number of student records. For each record, a number of attributes that represent a student's academic accomplishment are divided as follows:

*1) Student-Related Attributes*

These attributes highlight the status of the students, including:

- Module Mark: Student's mark in a certain module.
- Exam Mark: The mark achieved by a student on the exam-based assessment part.
- Cswk Mark: The mark achieved by a student on the coursework-based assessment part.

*2) Module-Related Attributes*

They describe a certain module and its characteristics. These attributes include:

- Module Code: An alphanumerical sequence of digits as a short code for each module.
- Exam Weighting (EXW): Based on the assessment method for each module, this attribute represents the total marks reserved for the exam-based assessment out of 100, i.e. it can be seen as the ratio of exam-based assessment to the total mark of a module.
- Coursework Weighting (CWW): Alternatively, this attribute indicates the ratio of coursework-based assessment to the total mark of a module. For each module, the values of these two attributes complement each other to reach the maximum total mark (100). For each department, different ratios between exam weighting and coursework weighting will be described in detail.
- Module Assessment Method: this categorical attribute represents the ratio between exam and coursework based assessment for a given module. This attribute may take the value of "Exam", Coursework", or "Both" where the "Exam" indicates that the Exam Weighting is 100 while the Coursework Weighting is 0, "Coursework" represents the opposite, and finally "Both" represents all ratios in between.

*3) Program-Related Attributes*

These attributes describe the program in which the student is registered, namely: Program Code, which represents the program in which a student is enrolled.

*B. Causes of Errors in Educational Data*

Prior to storing data in databases, data normally are processed by human interaction, computation or both. The sources of errors in databases are categorized into four main types: Data entry errors, measurement errors, distillation errors, and data integration errors [13]. As mentioned above, since educational data are usually not measured, the errors in them can be caused by humans (i.e. through data entry errors) so these data have minimum or zero noise compared to other, non-educational, data types. Alternatively, the data we have are normally referred to as quantitative data, which are defined as a set of integers or floating point numbers that measure quantities of interest [13].

*C. Missing Data Fields*

As mentioned above, educational data are characterized by minimum noise, or sometimes they can be noise-free. The data we have are not noise-free because of the fact that some values are missing. The missing values most of the times are left empty because they are implicitly shown in other fields. For example, the *Module Desc* field is barely filled and this creates an obstacle if this field is used by any statistical or data mining tool. Hence, a basic algorithm was developed that considers both the Exam Weighting and Coursework Weighting fields to fill the *Module Desc* categorical field as follows:

For Resit Desc in Records:
If Resit Desc= Exam or Resit Desc=Blanks:
If(Exam Weighting ==100) Assessment Method is through Exam only
Else If Resit Desc= Coursework or Resit Desc=Blanks
If(Coursework Weighting ==100)
Assessment Method is through Coursework only
Else If (Resit Desc=Both OR Resit Desc= Blanks
OR Resit Desc =Exam or Resit Desc= Coursework)
If (Exam Weighting !=0 AND Exam Weighting !=100)
Assessment Method is through both Exam and Coursework

There are other records that were removed from our datasets for the fact that they represent an extremely low percentage of the data which were detected through the method of outliner detection (i.e. readings that are in some sense "far" from what one would expect based on the rest of the data) [13]. For example, in the Department of Mechanical Engineering, the module mark of some students was not recorded because of the nature of the enrolled modules that either pass a student or not without including the pass or fail marks. The percentage of such students in the Department of Mechanical Engineering is 0.000047%, which can be neglected without affecting the remaining data.

*D. Most Significant Factors Affecting Students' Marks*

As mentioned upon investigating all the factors (attributes) mentioned above, some factors were immediately neglected as they were irrelevant to the issue at hand. On the other hand, some other factors are directly connected to the studied problem. Those factors were considered for cleaning and then for further processing in order to understand the effect of the assessment method on student academic accomplishments and how to achieve more accurate marks. *Module Mark, CWW, EXW, Module Code, Regno,* and *Module Desc* are the most significant factors that contribute in achieving more accurate marks. The choice of the above factors was based on observation and coefficient correlation. Those chosen based on observation are the ones that are directly connected to the problem at hand, such as *Module Mark, CW Weighting, EX Weighting, Regno*, and *Module Code*. In other words, the choice of the factors was based on the effect of those factors on the final outcome. For instance, a question to be asked is whether or not the *CW Weighting* can play a role in determining the accuracy of a student's mark based on the enrolled module's assessment method. This question can be directly and positively answered using Pearson correlation method. The method was applied on different factors to determine the correlation between them. Table II shows the coefficient correlation on selected factors where these values tend to be high for CW and EX to module mark and low to other factors:

TABLE II.     PEARSON CORRELATION OF DATA FACTORS

|  | EXW | CWW | CW mark | Exam mark |
|---|---|---|---|---|
| **EXW** |  |  |  |  |
| **CWW** | -1.00 |  |  |  |
| **CW mark** | 0.03 | -0.030 |  |  |
| **Exam mark** | 0..13 | -0.013 | 0.461 |  |
| **Module mark** | -0.126 | 0.126 | 0.690 | 0.966 |

By observing the correlation matrix, it can be noticed that Exam mark has the biggest positive effect on Module mark, followed directly by CW mark. On the other hand, some other factors have negative correlation to module mark, such as EX weighting. That is, the more the weight of the exam based assessment, the less the module mark will be. Therefore, this research has considered the most significant factors whose correlations values are the highest in order to investigate the accuracy of students' module marks under different assessment methods. The following section will present the pre-processed students' transcript data in a different way in order to increase the data mining accuracy by extracting a new factor from the assessment method and use it to refine student marks to ensure more accuracy prior to processing them using EDM.

IV. MODULE ASSESSMENT INDEX

In order to reflect the module assessment method on students' marks, the data have to be statistically investigated so that outliners can be identified, and information can be extracted and highlighted. The data have been processed through various steps, each of these steps either adds or removes data. The first step is to categorize the CW to EX ratios. In other words, to ease the processing of numerical EX and CW weighting fields, and combine their values into a categorical field that can later be a part of further processing, a combining and categorizing algorithm is proposed. The algorithm relies on the number of classes the ratio between CW to EX weightings can have. Namely, each department has its own classification of CW to EX weightings. Table III shows the different classes. By observing Table III, it is clear that no single department shares the same ratio classes with the other departments. Thus, filling the Table and therefore the data is not a solution since filling the empty cells in the Table results to filling the data tables with extra fields, yielding adding empty records which should be cleaned (i.e. removed) again. The solution would be to consider the department with the most number of classes to start with, and then generalize the findings on other departments while bearing in mind the change of ratios. The Table shows two departments that have 12 complete ratios (the CS and ECSEng departments), in this paper, the CS department was chosen for further processing.

TABLE III. CLASSES OF EX TO CW WEIGHTING RATIOS

| | Model Assessment Method | | | | | | | | | | | | | | | |
|---|---|---|---|---|---|---|---|---|---|---|---|---|---|---|---|---|
| **CW** | 0:100 | 25:75 | 30:70 | 34:66 | 35:65 | 40:60 | 45:55 | 50:50 | 55:45 | 60:40 | 65:35 | 70:30 | 75:25 | 80:20 | 85:15 | 90:10 | 100:0 |
| **Business** | ✓ | | | | ✓ | ✓ | | ✓ | | ✓ | | ✓ | ✓ | ✓ | | ✓ | ✓ |
| **CS** | ✓ | | ✓ | | ✓ | ✓ | ✓ | ✓ | | ✓ | | ✓ | ✓ | ✓ | | ✓ | ✓ |
| **CEng** | ✓ | ✓ | | | | ✓ | | ✓ | ✓ | ✓ | | ✓ | ✓ | ✓ | ✓ | | ✓ |
| **ECSEng** | ✓ | | | ✓ | | ✓ | | ✓ | | ✓ | ✓ | ✓ | ✓ | ✓ | ✓ | ✓ | ✓ |
| **Math** | ✓ | | ✓ | | | | | ✓ | | ✓ | | ✓ | ✓ | ✓ | | ✓ | ✓ |
| **MEng** | ✓ | ✓ | ✓ | | | | | ✓ | | ✓ | | ✓ | ✓ | ✓ | | ✓ | ✓ |

The algorithm first tries to ease the manipulation of ratio field by converting it into a categorical field. Figure 1 shows a simple procedure which will convert the EXW to categorical index values that can later be used in re-evaluating student marks. The output values are 0 to 11, representing the EXW to CWW ratio. For example, Module Assessment Index (MAI) value of 0 represents 0:100 EXW to CWW ratio, while MAI of 11 represents the opposite. The cases in the flowchart can be changed to different ratios from other departments in order to generalize this method.

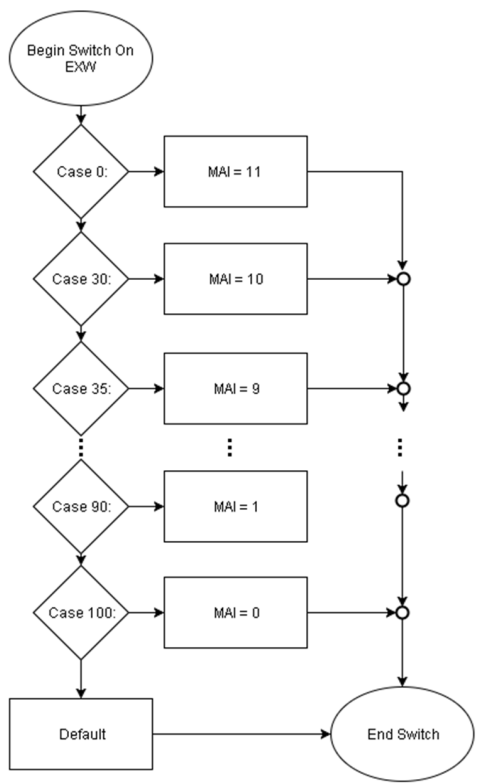

Fig. 1.  EXW to MAI matching – Department of CS

V. RE-EVALUATING STUDENTS MARKS BASED ON MAI

In order to uncover the relation between the student's module marks and MAI, simple quadratic regression was used. EXW and CWW were combined into one variable hence the simple quadratic regression is more suitable for one variable relation. The choice of quadratic over linear is based on the R-squared of the quadratic model (2.90%) which is higher than the one in the linear model (2.77%). By applying Simple Quadratic Regression on the data with MAI as a variable for module mark as a response, we achieved the following fitted regression line:

$$y = MM + 0.0035(MAI) - 0.05688(MAI)^2 \quad (1)$$

where $y$ is the re-evaluated module mark after fitting and MM is the current module mark.

As a next step the the data of CS Department are applied to (1) using the following pseudo code:

Input: Current Module Mark (MM) for each student and MAI
Output: Re-evaluated Module Mark (RMM)
For each MM:
If MAI=0 then RMM=MM
Else RMM=MM+0.0035(MAI) - 0.05688(MAI)^2

By applying the above pseudo code on student transcript data of the CS Department, an additional field will be added which contains the RMM for each student at the department. As a real example on applying the above, for the student with *regno* of *x* in the module "Server Side Programming" which has 0:100 *EXW* to *CWW* ratio (i.e. MAI=0), the RMM would be equal to MM, and since this student's MM is equal to 60, his RMM would be also 60. However, for the same student in another module, Social Informatics that has a *MAI* value of 5, the student's MM of 55 is re-evaluated as 57.03 after applying the formula. The new values of RMM are self-explained. Table I compares the average module marks for student attending modules with different assessment methods. That is, the more the percentage of *CWW* (yields more MAI), the more the added marks to MM, and vice versa. Regardless of the fact that this increase on marks has been proved in literature, yet it appears to be that none of the previous studies have considered this increase as a feedback to re-evaluate the marks. Therefore, it is vital to consider this feedback generation for the educational data mining pre-processing phase. This means that the formulation of the relation between MAI (which reflects the ration between EXW and CWW) and the current MM should be a part of the educational data mining processes. Figure 2 shows the additional processes on raw students' module marks. The goal is to take the differences of the assessment methods into account.

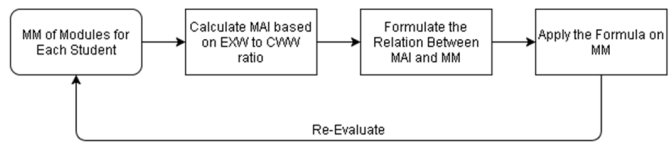

Fig. 2. Re-evaluating students' module marks

## VI. THE EFFECT OF MAI ON DM'S PREDICTION ACCURACY

In order to verify the effect of the newly constructed variable, the prediction of a student's second year average mark using only his first year's average compared to the predicted second year's average using first year's average and the newly constructed variable, MAI are compared. This comparison may highlight how accurate can the prediction process be by adding new attributes that reflect the nature of the module. Two prediction techniques were investigated: Orange Data Mining Tool: Naive Bayes [15, 16] and Random Forest [17]. The Random Forest showed more accuracy in terms of Classification Accuracy (CA) as shown in Table IV, hence it was used to evaluate the role of MAI in increasing prediction accuracy.

TABLE IV. CLASSIFICATION EVALUATION OF PREDICTION TECHNIQUES

| Method | AUC | CA | F1 | Precision | Recall |
|---|---|---|---|---|---|
| Random Forest | 0.827 | 0.942 | 0.194 | 0.273 | 0.15 |
| Naive Bayes | 0.876 | 0.924 | 0.330 | 0.281 | 0.40 |

Therefore, Random Forest was used to investigate the effect of including MAI as an attribute in prediction process. The data of 407 undergraduate students of the Computer Science Department, including their first and second years' average marks and their average MAI for both years were considered. The comparison has been done with and without the existence of MAI. Tables V and VI show the confusion matrices of comparisons and how MAI affected the accuracy of predicting the students' second year average marks. Table VI shows an enhancement on the prediction probabilities for each of the mark classes (Fail, Pass, Third, Lower second, Upper second, and First) compared to the probabilities of prediction shown in Table V. The Random Forest showed more accuracy in terms of CA when MAI was considered as shown in Figure 4 with an accuracy value 0.942, but when MAI was excluded, as shown in Figure 3, the accuracy was lower with a value of 0.874.

TABLE V. PREDICTING SECOND YEAR'S AVERAGE WITHOUT MAI

| | Fail | First | Lower second | Pass | Third | Upper second | ∑ |
|---|---|---|---|---|---|---|---|
| **Fail** | 57.1% | 2.9% | 3.9% | NA | 5.0% | 2.2% | 12 |
| **First** | 0.0% | 87.1% | 3.9% | NA | 0.0% | 6.6% | 72 |
| **Lower second** | 28.6% | 1.4% | 72.5% | NA | 10.0% | 7.3% | 52 |
| **Pass** | 0.0% | 0.0% | 0.0% | NA | 0.0% | 2.2% | 3 |
| **Third** | 14.3% | 1.4% | 2.0% | NA | 80.0% | 4.4% | 25 |
| **Upper second** | 0.0% | 7.1% | 17.6% | NA | 5.0% | 77.4% | 121 |
| **∑** | 7 | 70 | 51 | 0 | 20 | 137 | 285 |

TABLE VI. PREDICTING SECOND YEAR'S AVERAGE INCLUDING MAI AS AN ATTRIBUTE

| | Fail | First | Lower second | Pass | Third | Upper second | ∑ |
|---|---|---|---|---|---|---|---|
| **Fail** | 100.0% | 3.9% | 7.1% | 0.0% | 10.3% | 1.5% | 15 |
| **First** | 0.0% | 89.6% | 0.0% | 0.0% | 3.4% | 3.1% | 74 |
| **Lower second** | 0.0% | 2.6% | 85.7% | 0.0% | 0.0% | 5.3% | 45 |
| **Pass** | 0.0% | 0.0% | 0.0% | 100.0% | 0.0% | 0.0% | 2 |
| **Third** | 0.0% | 0.0% | 2.4% | 0.0% | 82.0% | 2.3% | 28 |
| **Upper second** | 0.0% | 3.9% | 4.8% | 0.0% | 3.4% | 87.8% | 121 |
| **∑** | 4 | 77 | 42 | 2 | 29 | 131 | 285 |

## VII. RESULTS AND DISCUSSION

By refining students' marks, they either increase or decrease depending on the ration between EXW to CWW for each student during his study. For instance, considering the student *x* as an example, who was enrolled in 32 modules during this study at the CS Department. Out of 32 modules, 19 are 100% exam-based assessed modules, 7 are assessed by a mixture of coursework and examination, while only 6 modules are 100% assessed by coursework only. Despite that the majority of the modules are assessed through examination only, which means that the student gets no extra marks compared to coursework-based modules, the rest of the modules give the student extra marks and hence add to his overall average. In numbers, 19 modules have 0 MAI value, which means that RMM = MM. On the other hand, the rest of the modules have values of MAI ranging from 1 to 11, which means that the RMM is always less than MM for those 13 courses. This decrement in the marks is due to the fact that students get higher marks in modules that are assessed by coursework or a mixture of coursework and examinations. Therefore, to balance the module marks and the overall average, the formula decreases the module marks by varying percentage depending on the EXW to CWW ratios. Table VII shows the differences in module marks and overall average for the example student.

TABLE VII. REFINING EXAMPLE STUDENT'S MARKS BASED ON ENROLLED MODULES' ASSESSMENT METHODS

|  | Average MM | Average RMM |
|---|---|---|
| 19 Exam-based modules | 48.6 | 48.6 |
| 6 Coursework-based modules | 60.3 | 53.5 |
| 7 Mix of EX and CW modules | 60.4 | 59.1 |
| Total of 32 Modules | 53.4 | 51.8 |

By following the procedure in Figure 2 on a real student's marks, Table VII shows that the RMM remained unchanged for the student when the assessment method of the enrolled modules was purely exam-based. Alternatively, when the assessment method of the enrolled modules was purely coursework-based, the RMM was refined down on an average mark of 6.8 compared to MM. Finally, a mixture of EX and CW based modules yields less refinement of the RMM (1.3 marks) compared to MM for the same student. This means that the student in the example who is taking 32 modules of different types may have his average marks refined down by 1.6 by the proposed procedure.

## VIII. CONCLUSIONS AND FUTURE WORK

During the last few decades, there has been an increased interest in coursework-based assessment in the UK and some other countries due to its various educational and personal advantages such as its learning effectiveness and the lack of time limits and stress. This increased interest had led to discovering that students who are assessed by coursework tend to achieve higher marks than those who are assessed by examinations in the same modules. However, to the best of our knowledge, no studies have considered this increase in marks in the data pre-processing phase. More generally, this led to a conclusion that applying conventional DM processes on Educational Data may not produce accurate results. In this paper, a model that refines students' marks based on enrolled modules' assessment methods has been proposed. The key to refine students' marks is to develop an index that categorizes ratios between exams to Coursework Weightings. This index – MAI – has been extracted using simple quadratic regression on the Module Marks (MM) variable. Based on MAI values, the Refined Module Marks (RMM) were calculated and compared to the original MM. The comparison showed that based on the percentage of pure exam-based assessed modules, the RMM may be less or more than the original MM. This study shows that the pre-processing phase of Educational Data should include additional sub-phases that deal not only with noise or missing data, but also with data refinement to cope with differences between various educational systems. The effect of MAI on the prediction process was also investigated using Random Forest classification. It was shown that considering MAI as attribute increases the accuracy of predicting students' second year averages based on their first year averages.

The findings of this paper will be generalized in different Departments and Universities that may have various assessment methods, in other words, the refinement procedure should be considered as a sub-process within EDM data pre-processing phase when dealing with different assessed modules and their marks.